# Dispersive Property of the Quantum Vacuum and the Speed of Light

H. Razmi [1], N. Baramzadeh [2] and H. Baramzadeh [3]

Department of Physics, The University of Qom, 3716146611, Qom, I. R. Iran.

[1] razmi@qom.ac.ir , razmiha@hotmail.com;  [2] nbaramzadeh@yahoo.com; [3] hbaramzadeh@yahoo.com

**Abstract**

We want to study the influence of the quantum vacuum on the light propagation. At first, by working in the standard linear quantum theory of the electromagnetic fields, it is shown that the electric permittivity and the magnetic permeability of the vacuum medium are changed; but, the resulting speed of light isn't modified. Then, taking into account nonlinear effects by considering the Euler-Heisenberg Lagrangian, the corresponding zero point (vacuum) energy and the resulting modification of the speed of light are found up to the first non-vanishing correction.

## 1. Introduction

In classical physics, the vacuum, as an empty space free of any energy and matter, has some physical properties as the electric permittivity $\varepsilon_0$ and the magnetic permeability $\mu_0$ where the finite value of the speed of light in vacuum is written in terms of them as $c = 1/\sqrt{\varepsilon_0 \mu_0}$. If the vacuum was "nothing", then it couldn't have nonzero electromagnetic properties with an impedance ($Z = \sqrt{\frac{\varepsilon_0}{\mu_0}}$) value of about 377 *Ohms* well-known in electrical and communication engineering [1]. In quantum physics, the zero point (vacuum) state has nontrivial properties. The spontaneous emission of radiation [2], the Lamb shift [3], the Casimir effect [4], and the radiation corrections [5] are explained based on the quantum vacuum fluctuations well-known in the standard quantum field theory. The quantum vacuum energy is considered in the investigation of some other famous subjects as the cosmological constant (the dark energy) problem [6]. In recent years, some researchers have tried to show that the experimentally verified values of the electromagnetic constants $\varepsilon_0$ and $\mu_0$ may be fundamentally found based on the properties of the quantum vacuum as a polarized medium [7-9]. The scenario considered in these researches is based on this idea that the quantum vacuum is a medium which is full of continuously created and annihilated virtual particles. These particles are named "virtual" not because they aren't real, but because they aren't observable due to the uncertainty principle constraint on their lifetime and energy values.

In this paper, we want to study the propagation of light in the vacuum based on a description of the quantum vacuum as an effective polarized medium. At first, by working in the standard linear quantum theory of the electromagnetic fields, it is shown that the electric permittivity $\varepsilon_0$ and the magnetic permeability $\mu_0$ are changed; but, the pre-defined value of the speed of light $c = 1/\sqrt{\varepsilon_0 \mu_0}$ remains unmodified. Then, by working with the Euler-Heisenberg Lagrangian [10] and considering its nonlinear effects up to the first non-vanishing correction, it is shown that the speed of light is modified due to the dispersive property of the quantum vacuum medium.

## 2. The Electromagnetic Properties of the Quantum Vacuum and the Speed of Light

In this section we want to see how the propagation of a real photon in the quantum vacuum influences the vacuum electric permittivity and magnetic permeability and then check if the speed of light changes.

### 2.1. A Fundamental Quantum Mechanical Analysis

Let us consider a free photon propagating with the frequency $\omega$ in a volume box of the scale length of the order of an electron Compton wavelength in which a virtual pair of electron-positron polarize the vacuum. In natural unit, the corresponding electric and magnetic fields $E$ and $B$ are simply written as:

$$E = B = \sqrt{\frac{\omega}{\lambda_c^3}} = \sqrt{\omega m_e^3} \qquad (1).$$

The corresponding electric polarization vector, by definition, is equal to the number of the electric dipole moments per unit volume:

$$P_0 = \frac{dp}{dV} = \frac{e\lambda_c}{\lambda_c^3} = em_e^2 \qquad (2).$$

Therefore, the vacuum electric permittivity is found as:

$$\tilde{\varepsilon}_0 = \frac{P_0}{E} = e\sqrt{\frac{m_e}{\omega}} \qquad (3).$$

Similar consideration for the magnetic permeability of the quantum vacuum due to the magnetic current loop of the virtual electron-positron pair leads us to the following expressions:

$$M_0 = \frac{dm_0}{dV} = \frac{IA}{\lambda_c^3} = \frac{e\,\lambda_c^2}{\tau\,\lambda_c^3} = em_e^2 \qquad (4).$$

$$\tilde{\mu}_0 = \frac{B}{M_0} = \frac{1}{e}\sqrt{\frac{\omega_e}{m_e}} \qquad (5).$$

where $A$ is the area of the current loop of the radius of the order of a fundamental electron Compton wavelength, $m_0$ and $M_0$ are the well-known magnetic dipole moment and the magnetization vector.

Using (3) and (5), it is found that the resulting speed of light remains unmodified:

$$v = \frac{1}{\sqrt{\tilde{\varepsilon}_0 \tilde{\mu}_0}} = 1(=c) \qquad (6).$$

**2.2 The Vacuum Polarization and the Modified Propagator**

As a simple quantum field theoretical development to the approach in [8], using the well-known modified photon propagator due to the vacuum polarization $D'_{F_{\mu\nu}}(k) = Z_3 D_{F_{\mu\nu}}(k)$ in which the coefficient $Z_3$ is the well-known renormalization constant [11], and knowing that the propagator of any field is a function of the square form of it, the electromagnetic field potential is renormalized as:

$$A_\mu^R(k) = \sqrt{Z_3}\, A_\mu^0(k) \qquad (7).$$

Application of the appropriate Fourier transform relations for the energy-momentum to the space-time leads us to the following modified forms of the electric potential, the magnetic potential, the electric field, the magnetic field, the electric polarization vector, and the magnetization vector:

$$\phi' = \sqrt{Z_3}\phi \qquad \vec{A}' = \sqrt{Z_3}\vec{A} \qquad \vec{E}' = \sqrt{Z_3}\vec{E} \qquad \vec{B}' = \sqrt{Z_3}\vec{B}$$

$$\vec{P}_{vacuum} = \varepsilon_0\sqrt{Z_3}\vec{E} \qquad \vec{M}_{vacuum} = \frac{\sqrt{Z_3}}{\mu_0}\vec{B} \qquad (8).$$

The direct result of the above relations is that the vacuum has been polarized with the following modified forms of the electric permittivity and the magnetic permeability:

$$\varepsilon_{vacuum} = \sqrt{Z_3}\varepsilon_0 \qquad \mu_{vacuum} = \frac{\mu_0}{\sqrt{Z_3}} \qquad (9).$$

Of course, the speed of light doesn't change:

$$v = \frac{1}{\sqrt{\varepsilon_{vacuum}\mu_{vacuum}}} = \frac{1}{\sqrt{\varepsilon_0\mu_0}} = c \qquad (10).$$

## 3. The Quantum Vacuum as a Nonlinear Dispersive Medium

In order to consider the nonlinear effects, it is appropriate to use the Euler-Heisenberg formulation of quantum electrodynamics. In what follows, after a short introduction to the effective Euler-Heisenberg theory, it is shown that the vacuum behaves as a dispersive medium and modifies the speed of light.

### 3.1. The Euler-Heisenberg Lagrangian and Hamiltonian Densities

As is well-known, the effective low energy (up to the second order of the fine structure constant) Euler-Heisenberg Lagrangian density is [10, 12]:

$$\mathcal{L}_{EH} = \frac{1}{2}(E^2 - B^2) + \frac{2\alpha^2}{45 m_e^4}\left[(E^2 - B^2)^2 + 7(E.B)^2\right] \quad (11).$$

In terms of the four potential $A_\mu$ and by applying the Lorentz gauge condition, the Euler-Heisenberg Lagrangian and Hamitonian densities are found as:

$$\mathcal{L}_{EH} = -\frac{1}{2}\left(\dot{A}_\mu \dot{A}^\mu - \vec{\nabla} A_\mu . \vec{\nabla} A^\mu\right) + \frac{\alpha^2}{90 m_e^4}\left\{4\left(\dot{A}_\mu \dot{A}^\mu - \vec{\nabla} A_\mu . \vec{\nabla} A^\mu\right)^2\right.$$

$$\left. -7\left(\dot{A}_\mu \dot{A}^\mu - \vec{\nabla} A_\mu . \vec{\nabla} A^\mu\right)\left(\dot{A}_\alpha \dot{A}^\alpha - \vec{\nabla} A_\alpha . \vec{\nabla} A^\alpha\right)\right\} ;$$

$$\mathcal{H}_{EH} = -\frac{1}{2}\left(\dot{A}_\mu \dot{A}^\mu + \vec{\nabla} A_\mu . \vec{\nabla} A^\mu\right) - \frac{\alpha^2}{30 m_e^4}\left\{\left(3\dot{A}_\mu \dot{A}^\mu + \vec{\nabla} A_\mu . \vec{\nabla} A^\mu\right)\left(\dot{A}_\alpha \dot{A}^\alpha - \vec{\nabla} A_\alpha . \vec{\nabla} A^\alpha\right)\right\} \quad (12).$$

The Hamiltonian function consists two terms where the first term is the standard well-known function:

$$H_1 = \int -\frac{1}{2}\left(\dot{A}_\mu \dot{A}^\mu + \vec{\nabla} A_\mu . \vec{\nabla} A^\mu\right) d^3x \quad (13),$$

and the second term is the nonlinear correction:

$$H_2 = -\frac{\alpha^2}{30 m_e^4}\int \left(3\dot{A}_\mu \dot{A}^\mu + \vec{\nabla} A_\mu . \vec{\nabla} A^\mu\right)\left(\dot{A}_\alpha \dot{A}^\alpha - \vec{\nabla} A_\alpha . \vec{\nabla} A^\alpha\right) d^3x \quad (14).$$

## 3.2. The Dispersion Relation due to the Nonlinear Effects

By considering the standard Fourier expansion[1] in the second quantization of the field $A_\mu(x)$ in terms of the creation and annihilation operators in a volume box $V$ [5], the energy expectation values $(E_1)_n$ and $(E_2)_n$ are simply calculated in terms of the photons number $n_r$ and their momentum $\vec{k}$ as in the following:

$$(E_1)_n = \langle n|H_1|n\rangle = \sum_{k,r} |\vec{k}| \left(n_r + \frac{1}{2}\right) ;$$

$$(E_2)_n = \langle n|H_2|n\rangle = -\frac{\alpha^2}{10 m_e^4} \frac{1}{4V} \sum_{k,k',r} (2n_r(n_r+1)+1) \left( 3|\vec{k}||\vec{k}'| - 2\vec{k}\vec{k}' - \frac{(\vec{k}.\vec{k}')^2}{|\vec{k}||\vec{k}'|} \right) \quad (15).$$

The first expression $(E_1)_n$ is the well-known zero point (vacuum) energy in the standard linear quantum electrodynamics. The second correction term $(E_2)_n$ is a more complicated expression which can be simplified by some considerations. The distance scale corresponding to the energy regime under consideration here is constrained up to the electron Compton wavelength; this constraint and this fact that the nonlinear perturbation term is at the second order of the fine structure constant ($\alpha^2$), lead us to apply a cutoff value for the momentum up to the electron mass $m_e$. Considering this cutoff value and using $\frac{1}{V}\sum_{\vec{k}'} \rightarrow \frac{1}{(2\pi)^3}\int d^3k'$ result in:

$$(E_2)_n = -\frac{\alpha^2}{120} \sum_{k,r} |\vec{k}|(2n_r(n_r+1)+1) \quad (16).$$

Using (15) and (16), the total corrected zero point (vacuum) energy is found as:

---

[1] Why can we use the standard Fourier expansion while we are considering nonlinear effects? In response to this question, we should say that the nonlinear corrections here are very small so that the electromagnetic field variation is enough smooth that we can work in the same way as in a perturbative linear regime theory.

$$E_0 = \sum_{k,r} \frac{1}{2}(1 - \frac{\alpha^2}{60})|\vec{k}| \qquad (17).$$

The above result reasons on this fact that the quantum vacuum is a dispersive medium with the following dispersion relation:

$$\omega(\vec{k}) = (1 - \frac{\alpha^2}{60})|\vec{k}| \qquad (18).$$

**3.3 Modification of the Speed of Light**

According to the dispersion relation (18), the speed of light should be modified as (coming back from the natural unit):

$$v = c(1 - \frac{\alpha^2}{60}) \qquad (19).$$

The numerical correction found in (19) is about one per million which seems it can be easily verified in experience; but, the main challenge is in the experimentally achievement of the required value for the electric or magnetic fields being able to polarize the vacuum [12].

**4. Discussion**

Why the result (19) is incompatible with the Lorentz invariance? Since the quantum vacuum responses to the motion of the light as an effective medium with iterative back reactions, dealing with variable value of the speed of light is unavoidable. Such a variable value for the speed of light is considered in a number of beyond standard models (e.g. different approaches to quantum gravity) in them the Lorentz invariance principle is violated. The swift short gamma ray bursts is an already known example for the energy-dependent velocity of light at smaller energies than the quantum gravity high energy scales [13].

5. Conclusion

In this paper, considering the quantum vacuum as an effective medium, it has been discussed about the origin of the electromagnetic properties of the vacuum based on the radiation corrections due to the vacuum polarization. It has been shown that although the electric permittivity and the magnetic permeability of the vacuum are changed due to the polarization made by the propagation of a real photon, the speed of light remains unchanged while one works in the standard linear theory of electrodynamics. By considering the nonlinear corrections in the Euler-Heisenberg model in an energy regime up to a value less than the electron mass, the dispersion relation (18) and the corresponding modification of the speed of light (19) have been found. Such quantum vacuum influences have been already known in a number of interesting effects as light by light scattering [14-15] and photon splitting in vacuum [16].

**References**

[1] P. C. Clemmow, *An Introduction to Electromagnetic Theory*, p. 183. Cambridge University Press (1973).

[2] P. A. M. Dirac, The Quantum Theory of the Emission and Absorption of Radiation, Proc. Roy. Soc. Lond. A.**114**, 243 (1927).

[3] H. A. Bethe, The Electromagnetic Shift of Energy Levels, Phys. Rev. **72**, 339 (1947).

[4] K. A. Milton, *The Casimir Effect: Physical Manifestations of Zero-Point Energy*, World Scientic (2001).

[5] F. Mandl and G. Shaw, *Quantum Field Theory*, p. 175, Wiley, New York (2010).

[6] S. Weinberg, The Cosmological Constant Problem, Rev. Mod. Phys. **61**, 1 (1989).

[7] M. Urban , F. Couchot , X. Sarazin and A. Djannati-Atai, The quantum vacuum as the origin of the speed of light, Eur. Phys. J. D **58**, 673 (2013).

[8] G. Leuchs, A. S. Villar and L. L. Sanchez-Soto, The quantum vacuum at the foundations of classical electrodynamics, Appl. Phys. B **100**, 9 (2010).

[9] G. Leuchs and L. L. Sanchez-Soto, A sum rule for charged elementary particles, Eur. Phys. J. D (2013).

[10] W. Heisenberg and H. Euler, Consequences of Dirac Theory of Positron, Zeit. f. Phys. **98**, 714, (1936).

[11] W. Greiner and J. Reinhart, *Quantum Electrodynamics*, Springer, NewYork (2003).

[12] R. Battesti and C. Rizzo, Magnetic and electric properties of a quantum vacuum, Rep. Prog. Phys. **76**, 016401 (2013).

[13] M. G. Bernardini, et al, Limits on quantum gravity effects from Swift short gamma-ray bursts, A&A 607, A121 (2017).

[14] O. Halpern, Scattering Processes Produced by Electrons in Negative Energy States, Phys. Rev. 44, 855 (1933).

[15] G. Breit and J. A. Wheeler, Collision of Two Light Quanta, Phys. Rev. **46**, 1087 (1934).

[16] Z. Bialynicka-Birula and I. Bialynicki-Birula, Nonlinear Effects in Quantum Electrodynamics: Photon Propagation and Photon Splitting in an External Field, Phys. Rev. D **2**, 2341 (1970).